\documentclass[amsmath,amssymb,widetext,10pt,a4paper]{revtex4}
\usepackage{graphicx}

\begin{document}
\title{The resonating valence bond wave functions in quantum antiferromagnets}
\author{Alberto Parola}
\affiliation{Dipartimento di Fisica e Matematica, Universit\`a dell'Insubria, I-22100 Como, Italy}
\author{Sandro Sorella}
\author{Federico Becca}
\affiliation{INFM-Democritos, National Simulation Centre, and SISSA, I-34014 Trieste, Italy}
\author{Luca Capriotti}
\affiliation{Credit Suisse First Boston (Europe) Ltd. One Cabot Square, London E14 4QJ, United Kingdom}


\maketitle
Projected-BCS wave functions have been proposed as the paradigm 
for the understanding of disordered spin states (spin liquids). 
Here we investigate the properties of these wave functions showing how 
Luttinger liquids, dimerized states, and gapped spin liquids may be described 
by the same class of wave functions, which, therefore, represent an extremely 
flexible variational tool. A close connection between spin liquids and 
``frozen'' superconductors emerges from this investigation.   

\vspace{15mm}
Many years after the first proposal~\cite{anderson},
the very existence of a spin-liquid ground state in two-dimensional (2D)
spin-$1/2$ models is still a controversial issue.
Short-range resonating-valence-bond (RVB)~\cite{figuerido} phases
with exponentially decaying spin-spin correlations 
and no broken lattice symmetry (i.e., with no valence-bond order),
are conjectured to be stabilized by quantum fluctuations.
However, while it is possible to show that spin-liquid ground states
can be found in quantum dimer models~\cite{qdm}, the numerical evidence in
favor of disordered ground states in Heisenberg-like antiferromagnets
is still preliminary~\cite{rainbow,hex,kag2,mse}.
Renewed interest on this topic is triggered by the possible realization 
of a frustrated Bose-Hubbard model with tunable interactions, by 
trapping cold atomic clouds in optical lattices~\cite{tosi}. 
A thorough analysis of the zero-temperature phase diagram of
these models will probably require an outstanding numerical effort,
however, it is possible to tackle the spin-liquid problem by a different,
less ambitious but remarkably informative, perspective.
Here, we present a detailed analysis of the properties of a class of RVB 
wave functions, which have been shown to represent extremely good variational 
states for a wide class of microscopic 
Hamiltonians, ranging from effective low-energy models for correlated electrons
on the lattice to realistic models of atoms and 
molecules~\cite{rainbow,dagotto,casula}. In particular, we will
show that both spin-liquid states and valence-bond crystals 
may result from the ``freezing'' of Cooper pairs in the zero-doping limit 
of correlated electron models.

Following Anderson's suggestion~\cite{anderson}, we define the class of
projected-BCS (pBCS) wave functions in a $N$-site spin lattice,
starting from the ground state of a suitable BCS Hamiltonian:  
\begin{eqnarray}
H(t,\Delta)&=&\sum_{i,j \sigma} (t_{ij}-\mu\,\delta_{ij})\, c^\dagger_{i,\sigma}c_{j,\sigma}
-\sum_{i,j} \left [ \Delta_{ij} c^\dagger_{i,\uparrow}c^\dagger_{j,\downarrow} +
\Delta_{ij}^* c_{j,\downarrow}c_{i,\uparrow} \right ] 
\nonumber\\
&=&\sum_{k \sigma} (\epsilon_{k}-\mu)\, c^\dagger_{k,\sigma}c_{k,\sigma}
-\sum_{k} \left [ \Delta_{k} c^\dagger_{k,\uparrow}c^\dagger_{-k,\downarrow} +
\Delta_{k}^* c_{-k,\downarrow}c_{k,\uparrow} \right ],
\label{hbcs}
\end{eqnarray}
where the bare electron band $\epsilon_k$ is real and both $\epsilon_k$ 
and $\Delta_k$ are even functions of $k$. A chemical potential $\mu$ is
introduced in order to fix the number of electron equal to the number of sites. 
In order to obtain a class of non-magnetic, translationally invariant, 
singlet wave functions for spin-1/2 models, the ground state 
$|BCS \rangle$ of Hamiltonian~(\ref{hbcs}) is then restricted to the physical 
Hilbert space of singly-occupied sites by the Gutzwiller projector $P_G$.

The first feature of such a wave function we want to discuss is the 
{\it redundancy} implied by the electronic representation of a spin state, 
i.e., the extra symmetries which appear when we write a spin state as
Gutzwiller projection of a fermionic state.
In turn, this property reflects in the presence of
a local, i.e., gauge, symmetry of the fermionic problem, as already
pointed out several years ago~\cite{su2}. Let us consider a generic
spin operator defined on a site in fermionic representation:
$X_{\alpha\beta}=c^\dag_\alpha c_\beta$.
This operator acts in the Hilbert subspace of singly-occupied sites (which
is left invariant under any spin Hamiltonian). It is easy to check that
the three SU(2) generators $N_0=(c^\dag_\uparrow c_\uparrow +
c^\dag_\downarrow c_\downarrow -1)/2$,
$N_+=c^\dag_\uparrow c^\dag_\downarrow$, $N_-=(N_+)^\dag$
commute, in the singly occupied site subspace, with $X_{\alpha\beta}$.
This property reflects the invariance of the operator $X_{\alpha\beta}$
under the usual $U(1)$ gauge transformation $g_\phi$:
\begin{equation}
g_\phi:\qquad c^\dag_\alpha \to e^{i\phi} c^\dag_\alpha,
\label{u1}
\end{equation}
and also by the SU(2) rotation $\Sigma_\theta$:
\begin{equation}
\Sigma_\theta: \qquad
\left (\begin{array}{c} c^\dag_\uparrow \cr c_\downarrow \end{array}\right )
\to \left ( \begin{array}{cc} \cos\theta & -i \sin\theta \cr
-i\sin\theta &\cos\theta \end{array} \right )
\left ( \begin{array}{c} c^\dag_\uparrow \cr c_\downarrow \end{array}\right ).
\label{sigma}
\end{equation}
These transformations are local, i.e., can be performed on each
site independently leaving every spin state invariant: they
generate the SU(2) gauge symmetry group.

Let us now consider the Gutzwiller projected BCS state
in a Heisenberg-like model on a lattice with an even number $N=2n$ of sites.
\begin{equation}
|pBCS\rangle = P_G |BCS\rangle = P_G 
\prod_k (u_k+v_k c^\dagger_{k,\uparrow}c^\dagger_{-k,\downarrow}) |0\rangle,
\label{bcs2}
\end{equation}
where the product is over all the $N$ wave vectors in the Brillouin zone.
The diagonalization of Hamiltonian~(\ref{hbcs}) gives explicitly
$$
u_k = \sqrt{{ E_k+\epsilon_k\over 2 E_k} }  \qquad\qquad
v_k = {\Delta_k\over |\Delta_k |} \sqrt{{ E_k-
\epsilon_k\over 2 E_k} }  \qquad\qquad
E_k = \sqrt{ \epsilon_k^2 +|\Delta_k|^2 } \qquad\qquad ,
$$
while the BCS pairing function $f_k$ is given by:
\begin{equation}
f_k = {v_k\over u_k} = {\Delta_k \over \epsilon_k + E_k}.
\label{fk}
\end{equation}
Clearly, the local gauge transformations previously defined 
change the BCS Hamiltonian, breaking in general the translation
invariance. In the following, we will restrict to the class of
transformations which preserve the translational symmetry of the lattice
in the BCS Hamiltonian, i.e., the subgroup of {\it global} symmetries
corresponding to site independent angles $(\phi,\theta)$. By applying the 
transformations~(\ref{u1}) and~(\ref{sigma}), the BCS Hamiltonian keeps 
the same form with modified couplings:
\begin{eqnarray}
t_{ij}&\to& t_{ij} \nonumber\\
\Delta_{ij}&\to&\Delta_{ij} e^{2i\phi}
\label{phi}
\end{eqnarray}
for $g_\phi$, while the transformation $\Sigma_\theta$ gives:
\begin{eqnarray}
t_{ij}&\to& \cos 2\theta \, t_{ij} -i\sin \theta\cos\theta \, (\Delta_{ij} -\Delta^*_{ij}) \nonumber\\
&=&\cos 2\theta \, t_{ij}+\sin 2\theta \, {\rm Im}\Delta_{ij} \nonumber\\
\Delta_{ij}&\to& (\cos^2 \theta \, \Delta_{ij} + \sin^2\theta \, \Delta^*_{ij})- i\sin 2\theta \, t_{ij}
\nonumber\\
&=&{\rm Re} \Delta_{ij} + i\left ( \cos 2\theta \, {\rm Im}\Delta_{ij}-\sin 2\theta \, t_{ij} \right ).
\label{theta}
\end{eqnarray}
These relations are linear in $t_{ij}$ and $\Delta_{ij}$ and, therefore, 
equally hold for the Fourier components $\epsilon_k$ and $\Delta_k$. 
Note that, being $\Delta_r$ an even function, the real (imaginary) part of 
its Fourier transform $\Delta_k$ equals the Fourier transform of the real 
(imaginary) part of $\Delta_r$.
It is easy to see that these two transformations generate the full $O(3)$ 
rotation group on the vector whose components are 
$(\epsilon_{k}, \,{\rm Re} \Delta_{k},\,{\rm Im}\Delta_{k})$.
As a consequence, the length $E_k$ of this vector is conserved by the full 
group. In summary, this shows that there is an infinite number of different
translationally invariant BCS Hamiltonians which, after projection, give the
same spin state. Choosing a specific representation does not affect the 
physics of the state but changes the pairing function $f_k$ before projection.
Within this class of states the only scalar under rotations, which can be 
given some physical meaning, is the BCS energy spectrum $E_k$.
Clearly, the projection operator will modify the excitation spectrum
associated to the BCS wave function. Nevertheless, the invariance
with respect to SU(2) transformations suggests that $E_k$ may 
reflect the nature of the physical excitation spectrum.

Remarkably, in one dimension it is easy to prove that such a class of wave 
functions is able to faithfully represent both the physics of Luttinger 
liquids, appropriate for the nearest-neighbor Heisenberg model, and the 
gapped spin-Peierls state, which is stabilized for sufficiently strong 
frustration. In fact, it is known~\cite{haldane} that the simple choice 
of nearest-neighbor hopping $t_{ij}$ ($\epsilon_k=-2t\,\cos k$, $\mu=0$ ) 
and vanishing gap function $\Delta_{ij}$ reproduces the exact solution of the 
Haldane-Shastry model, while choosing a next-nearest neighbor 
hopping ($\epsilon_k=-2t\,\cos 2k$, $\mu=0$) and a sizable nearest-neighbor 
pairing ($\Delta=4\sqrt{2}t\,\cos k$) we recover the Majumdar-Gosh 
state~\cite{maj}, i.e., the exact ground state of the frustrated Heisenberg 
model when the next-nearest-neighbor exchange constant $J_2$ is half of the 
nearest-neighbor one $J_1$. Note that in this case, the 
BCS dispersion $E_k$ is strictly positive, i.e., the BCS Hamiltonian is gapped. 

A quantitative analysis of this class of wave functions can be only carried 
out numerically. As an example, in Fig.1 we show that the spin-spin 
correlations $G(r)$ have a long-range tail, i.e., $G(r) \sim 1/r$, 
for both the known 
$\Delta=0$ solution~\cite{haldane} but also for our pBCS state with 
nearest-neighbor hopping and nearest- and third-neighbor $\Delta$, 
whose energy is indeed remarkably accurate~\cite{dagotto}. 
Moreover, Fig.2 shows that the same class of wave functions is able to 
display a clear dimer ordering as soon as the BCS dispersion $E_k$ shows a gap 
at the Fermi level. The agreement with the exact results (given by the
Density-Matrix Renormalization Group method) for the frustrated Heisenberg 
model at $J_2/J_1=0.4$ is excellent also in this case. 

\begin{figure}[!ht]
\includegraphics[width=8cm]{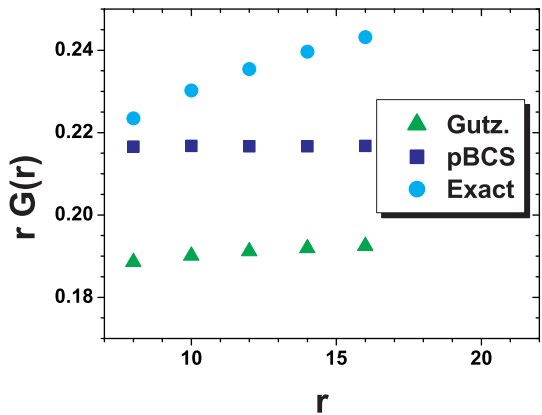}
\caption{Spin-spin correlation function at the largest distance:
$G(r)=(-1)^r\,\langle S_r^zS_1^z \rangle$ in a chain with L=2r sites as a 
function of $r$.
Circles: exact results for the Heisenberg model, Squares: pBCS wave function 
with nearest-neighbor hopping and third-neighbor $\Delta$. 
Triangles: Gutzwiller projected Fermi sea. 
A $1/r$ decay is clearly present in the variational wave functions. 
The exact result shows logarithmic corrections as predicted by conformal 
field theory.}
\end{figure}

\begin{figure}[!ht]
\includegraphics[width=8cm]{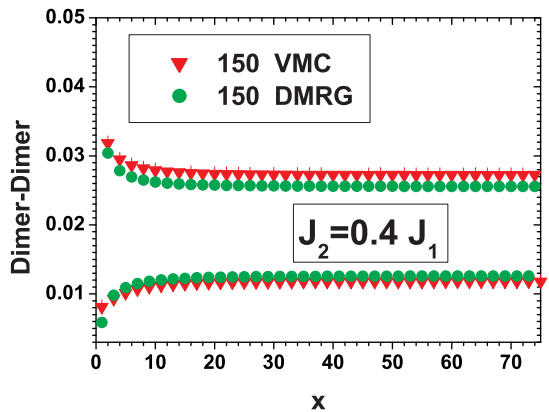}
\caption{Dimer-dimer correlation function 
$\langle S_{x+1}^z S_x^z S_1^zS_0^z \rangle$ 
as a function of $x$ calculated via the best variational pBCS function 
compared to the exact results obtained via Density-Matrix Renormalization 
Group (DMRG) in a $150$-site chain.}
\end{figure}

Now we specialize to the 2D square lattice and we
investigate whether it is possible to further 
exploit the redundancy of the fermion representation of a 
spin state in order to define a pairing function which,
before projection, breaks the reflection symmetries of the
lattice, while the projected state retains all the 
correct quantum numbers. We will show that, if suitable conditions are
satisfied, a fully symmetric projected BCS state is
obtained from a BCS Hamiltonian with fewer symmetries than
the original spin problem.

We first introduce a set of unitary operators.
\begin{itemize}
\item
{Spatial symmetries: $R_x (x,y)=(x,-y)$;
$R_{xy}(x,y)=(y,x)$. We define the transformation law of
creation operators $R\,c^\dagger_{j,\sigma}R^{-1}=c^\dagger_{Rj,\sigma}$
and the action of the operator on the vacuum $R\,|0\rangle=|0\rangle$.}
Note that these operators map each sublattice into itself.

\item
{ Particle-hole: $P_h c^\dagger_{j,\sigma} P_h^{-1}
= i\,(-1)^j c_{j,-\sigma}$, while the action of the $P_h$ operator 
on the vacuum state is  
$P_h|0\rangle=\prod_j c^\dagger_{j,\uparrow} c^\dagger_{j,\downarrow} |0\rangle$.}

\item
{ Gauge transformation: $G\,c^\dagger_{j,\sigma}G^{-1}=i\,c^\dagger_{j,\sigma}$
and $G\,|0\rangle=|0\rangle$.}

\end{itemize}

Clearly, $R_x$ and $R_{xy}$ are symmetries of the physical
problem (e.g., the Heisenberg model). $G$ is a symmetry because the
physical Hamiltonian has a definite number of electrons and
$P_h$ leaves invariant every configuration where each site is
singly occupied if the total magnetization vanishes: $N_\downarrow=
N_\uparrow=n$. Therefore, the previously defined operators commute
with the Heisenberg Hamiltonian and commute with each other
because reflections do not interchange the two sublattices.
The ground state of the Heisenberg model on a finite
lattice, if it is unique, must be simultaneous eigenstate
of all the symmetry operators. We want to investigate
sufficient conditions which guarantee that the projected
BCS state is indeed eigenstate of all these symmetries.

Let us consider a hopping term which just couples the two
sublattices: $\epsilon_{k+Q}=-\epsilon_k$ and
a gap function $\Delta=\Delta_s+\Delta_{x^2-y^2}+
\Delta_{xy}$ with contributions from different symmetries.
Moreover, we consider the case in which $\Delta_s$ and $\Delta_{x^2-y^2}$ 
will couple different sublattices while $\Delta_{xy}$ is restricted to the 
same sublattice. If this is the case, the BCS Hamiltonian
$H(t,\Delta_s,\Delta_{x^2-y^2},\Delta_{xy})$ transforms in the
following way under the different unitary operators:

\begin{eqnarray}
R_x H(t,\Delta_s,\Delta_{x^2-y^2},\Delta_{xy}) R_x^{-1} &=&
H(t,\Delta_s,\Delta_{x^2-y^2},-\Delta_{xy}) \nonumber \\
R_{xy} H(t,\Delta_s,\Delta_{x^2-y^2},\Delta_{xy}) R_{xy}^{-1} &=&
H(t,\Delta_s,-\Delta_{x^2-y^2},\Delta_{xy}) \nonumber \\
P_h H(t,\Delta_s,\Delta_{x^2-y^2},\Delta_{xy}) P_h^{-1} &=&
H(t,\Delta_s^*,\Delta_{x^2-y^2}^*,-\Delta_{xy}^*) \nonumber \\
G H(t,\Delta_s,\Delta_{x^2-y^2},\Delta_{xy}) G^{-1} &=&
H(t,-\Delta_s,-\Delta_{x^2-y^2},-\Delta_{xy}). \nonumber
\end{eqnarray}

Starting from these transformations, it is easy to define
suitable composite symmetry operators which indeed leave
the BCS Hamiltonian invariant. For instance,
let us consider the case in which $\Delta$ is real:
in this case we select the $R_xP_h$ and
$R_{xy}$ if $\Delta_{x^2-y^2}=0$ or $R_{xy}P_h G$ if $\Delta_s=0$.
We cannot set simultaneously $\Delta_{x^2-y^2}$ and $\Delta_s$
different form zero.
The eigenstates $|BCS\rangle$ of Eq.~(\ref{hbcs}) will be generally 
simultaneous eigenstates of these two composite symmetry operators with 
given quantum numbers, say $\alpha_x$ and $\alpha_{xy}$. Let us consider 
the effect of projection over these states:
\begin{eqnarray}
\alpha_x\, P_G |BCS\rangle &=& P_G R_xP_h |BCS\rangle 
= R_xP_h P_G |BCS\rangle = 
\nonumber \\
&=& R_x P_G |BCS\rangle,
\label{eigenx}
\end{eqnarray}
where we have used that both $R_x$ and $P_h$ commute with the projector
and that $P_h$ acts as the identity on singly occupied states.
Analogously, when a $d_{x^2-y^2}$ gap is present,
\begin{eqnarray}
\alpha_{xy}\, P_G |BCS\rangle &=& P_G R_{xy}P_h G|BCS\rangle =
R_{xy}P_h G P_G |BCS\rangle = \nonumber \\
&=&(-1)^{n} R_{xy} P_G |BCS\rangle.
\label{eigenxy}
\end{eqnarray}

These equations show that the projected BCS state with
both $xy$ and $x^2-y^2$ contributions to the gap has definite
symmetry under reflections, besides being translationally
invariant. The corresponding eigenvalues, for $n=N/2$ even,
coincide with the eigenvalues of the modified symmetry operators
$R_xP_h$ and $R_{xy}P_h G$ on the pure BCS state.
Note that the quantum numbers we have defined, refer to
the fermionic representation.
It turns out that an extremely good variational wave function
for the frustrated two dimensional Heisenberg antiferromagnet 
can be obtained by including gap functions of different 
symmetry~\cite{rainbow}.

An alternative, but equally interesting representation of the
pBCS state can be given in terms of Slater determinants through the 
following argument. 
The pBCS wave function can be written in real space as:
\begin{equation}
|pBCS\rangle = P_G \left [ \sum_{R<X} \sum_{\sigma}
f(\sigma,R;-\sigma,X)
c^\dagger_{R,\sigma} c^\dagger_{X,-\sigma} \right ]^n |0\rangle
\label{bcs}
\end{equation}
where $R$ and $X$ run over the lattice sites and $\sigma$ is
the spin index. The antisymmetric pairing function 
$f(\sigma,R;-\sigma,X)=-f(-\sigma,X;\sigma,R)$ is simply related to the 
Fourier transform $f(r)$ of the previously defined $f_k$ of Eq.~(\ref{fk}): 
$f(\uparrow,R;\downarrow,X)=f(R-X)$. In order to avoid double counting, 
a given, arbitrary, ordering of the lattice sites has been assumed in 
Eq.~(\ref{bcs}).
By expanding Eq.~(\ref{bcs}) and defining $f(\sigma,R;\sigma,X)=0$ we get
\begin{eqnarray}
|pBCS\rangle\,\, = \,\sum_{\begin{array}{c} R_1,\sigma_1\cdots R_n,\sigma_n \cr 
X_1,\sigma_1^\prime \cdots X_n,\sigma_n^\prime \end{array}} 
&f(\sigma_1,R_1;\sigma_1^\prime,X_1) \cdots f(\sigma_n,R_n;\sigma_n^\prime,X_n)&\,\,\times\nonumber \\
&|\sigma_1,R_1;\sigma_1^\prime,X_1\cdots \sigma_n,R_n;\sigma_n^\prime,X_n\rangle&
\label{mostro}
\end{eqnarray}
where the $2n$ labels ($R_1\cdots R_n;X_1\cdots X_n$) define a generic
partition of the $N=2n$ sites of the lattice.
We now rearrange the creation operators in the many body state of 
Eq.~(\ref{mostro})
$$
|\sigma_1,R_1;\sigma_1^\prime,X_1\cdots \sigma_n,R_n;\sigma_n^\prime,X_n\rangle=
c^\dagger_{R_1,\sigma_1} c^\dagger_{X_1,\sigma_1^\prime}\cdots
c^\dagger_{R_n,\sigma_n} c^\dagger_{X_n,\sigma_n^\prime} |0\rangle,
$$
according to a given {\it site} ordering in the lattice (irrespective of the 
spin).
This operation gives the ($S_z=0$) spin state $|\sigma_1\cdots \sigma_N\rangle$
multiplied by a phase factor equal to the sign $\epsilon$ of the associated
permutation. However, many distinct terms correspond to the same
spin state because they differ only by the way pairs are coupled:
$$
|pBCS\rangle = \sum_{\{\sigma_i\}}\sum_P \epsilon_P f(\sigma_1,R_1;\sigma_1^\prime,X_1)
\cdots f(\sigma_n,R_n;\sigma_n^\prime,X_n)
\,\,|\sigma_1\cdots \sigma_N\rangle
$$
Here $P$ runs over all the possible $(N-1)!!$ partitions of the $N=2n$ sites
of the lattices into pairs, for a given spin state 
$|\sigma_1\cdots \sigma_N\rangle$ uniquely identified by the variables 
$\{\sigma_i\}$.
The weight of the spin configuration
is exactly the Pfaffian of the $2n\times 2n$ antisymmetric matrix~\cite{pfaff}:
\begin{equation}
{\bf A} \,=\,\left ( \begin{array}{cc} \Big [ f(\uparrow,R_\alpha;\uparrow,R_\beta) \Big ] &
\Big [ f(\uparrow,R_\alpha;\downarrow,X_\beta)\Big ]\cr
\Big [ f(\downarrow,X_\alpha;\uparrow,R_\beta)\Big ] &
\Big [f(\downarrow,X_\alpha;\downarrow,X_\beta)\Big ]\end{array}\right ),
\label{matri}
\end{equation}
where the matrix has been written in terms of $n\times n$ blocks and $R_\alpha$
are the positions of the up spins in the $|\sigma_1\cdots \sigma_N\rangle$ 
state while $X_\alpha$ are the positions of the down spins.
The known relation
\begin{equation}
\left [ {\rm Pf} {\bf A} \right ]^2 = {\rm det} {\bf A},
\end{equation}
together with the fact that we considered $f(\sigma,R;\sigma,X)=0$
imply that ${\rm Pf} {\bf A} = {\rm det} \Big [ f(R_\alpha-X_\beta)\Big ]$~\cite{note}.
Therefore, the weight of the spin state $|\sigma_1\cdots \sigma_N\rangle$ 
may be written as the determinant of a matrix which depends on the BCS 
pairing function.

When we want to represent a spin state in a 2D lattice 
by the pBCS wave function, special care must be given to the definition 
of the phases of the pairing function $f(r)$.
As an example, let us consider a nearest-neighbor pairing function, i.e., 
we choose $|f(r)|=1$ when $r$ connects nearest neighbor sites in the lattice 
and zero otherwise. 
From Eq.~(\ref{bcs}) it is clear that the pBCS state is written as the sum 
of all possible partitions of the $N$-site lattice into singlets 
$(R_i,X_i)$ and the amplitude of a given partition is provided by the generic 
term of the Pfaffian of the matrix ${\bf A}$:
\begin{equation}
\epsilon_P\,f(R_1-X_1) \cdots f(R_n-X_n).
\label{sign}
\end{equation}
Here we introduced the usual convention to orient a singlet from sublattice 
$A$ to sublattice $B$ and we fix $R_i\in A$ ($X_i\in B$), while the 
permutation $P$ relates the chosen ordering of sites in the lattice and the 
sequence $(R_1,X_1\cdots R_n,X_n)$. Remarkably, as proved by 
Kasteleyn~\cite{pfaff}, in {\it planar} lattices it is possible to choose 
the phase of the function $f(r)$ so that the products of the form 
Eq.~(\ref{sign}) have all the same sign. In such a case, the
pBCS wave function exactly reproduces the short-range RVB state: 
the equal-amplitude superposition of all possible partitions of the lattice 
into nearest-neighbor singlets~\cite{figuerido}. 
In particular, on a rectangular lattice, the resulting 
pairing function has $s+id$ symmetry~\cite{read}, i.e., $f(r)=1$ on 
horizontal bonds and $f(r)=i$ on vertical bonds. In turn, this may be 
obtained from the BCS Hamiltonian~(\ref{hbcs})
by considering third neighbor hopping 
$\epsilon_k=-2t\,(\cos 2k_x + \cos 2k_y)$ (with $\mu=0$) and complex 
nearest-neighbor gap function $\Delta_k=8t\,(\cos k_x + i\cos k_y)$.

The properties of of the pBCS wave functions have been investigated by 
Lanczos technique and variational Monte Carlo method in the frustrated 
Heisenberg model on a square lattice, defined by the Hamiltonian:
\begin{equation}
H=J_1\sum_{\langle i,j\rangle} {\bf S}_i \cdot {\bf S}_j + 
J_2\sum_{\langle \langle i,j\rangle \rangle} {\bf S}_i \cdot {\bf S}_j,
\label{j1j2}
\end{equation}
where the first sum runs over the nearest neighbors and the second on the 
next-nearest neighbors (i.e., along the diagonal). When both coupling 
constants are positive the model is frustrated. 
In Fig.3 we compare the previously introduced dimer-dimer correlations 
$\langle S_j^z S_i^z S_1^zS_0^z \rangle$
(where $(i,j)$ and $(0,1)$ are nearest neighbor sites) obtained via the 
optimized pBCS wave function and the exact Lanczos results for the 
Hamiltonian~(\ref{j1j2}) at $J_2/J_1=0.55$.
The figure shows that the pBCS wave function is able to capture the 
correct behavior of correlations also in 2D.
In Fig.4 we report the size scaling of the squared dimer order parameter 
obtained via the optimal variational wave function in the pBCS class.
The absence of dimer order is clearly suggested by the variational approach, 
pointing toward the existence of a spin-liquid in the 2D $J_1{-}J_2$ model.  

\begin{figure}[!ht]
\includegraphics[width=8cm]{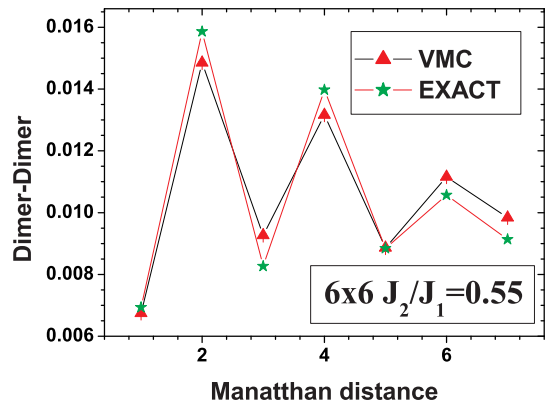}
\caption{Dimer-dimer correlations in the $6\times 6$ square lattice for the 
Hamiltonian~(\ref{j1j2}) at $J_2/J_1=0.55$. Stars: exact Lanczos results. 
Triangles: optimized pBCS wave function.} 
\end{figure}

\begin{figure}[!ht]
\includegraphics[width=8cm]{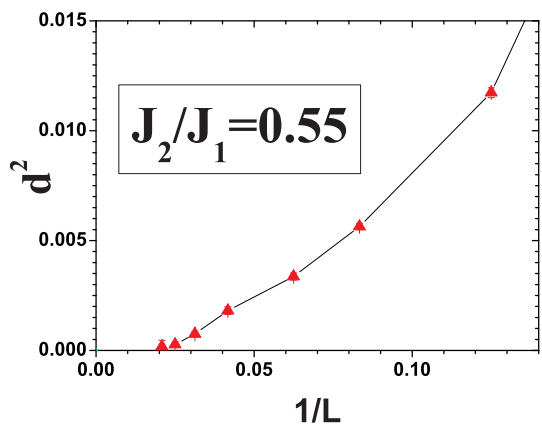}
\caption{Size scaling of the squared dimer order parameter as predicted by 
the pBCS wave function.}
\end{figure}

As a final remark, we like to comment on the fate of the insulating state
described by a pBCS wave function upon doping. When a limited number of
holes are injected into the lattice, it is likely that the basic structure 
of the pBCS state is not affected by the presence of mobile charges, 
being largely determined by the super-exchange interaction among spins. 
As long as the BCS pairs present in the insulating state remain well 
defined even at low doping, the system is generally expected to display 
superconducting properties. This possibility has been
explicitly verified in the so-called $t{-}J$ model~\cite{dagotto}, 
where the natural generalization of a pBCS wave function has been shown 
to give rise to off-diagonal long-range order.  
Real materials, like high-temperature superconductors, display a considerably 
richer physics and other effects may inhibit the actual realization of this 
scenario: charge-density waves may occur at special values of the doping
or the gain in hole kinetic energy may induce a global change in the spin 
state causing the breaking of electron 
pairs. Only a detailed study of the microscopic Hamiltonian will discriminate 
among these and other possibilities, and no general 
statement can be drawn on the basis of our analysis. However, 
our study suggests that a spin liquid (or even a valence bond crystal) may be 
thought of as an insulating state adiabatically connected to a superconducting 
phase which directly originates as soon as electrons are removed: 
this is a remarkable manifestation of the effects of electron correlations 
present in a gapped spin state which goes beyond standard band theory and the 
Fermi Liquid approach.

Partial support has been provided my MIUR through a PRIN grant. F.B. is
supported by INFM.



\begin{thebibliography}{99}
\bibitem{note} The ambiguity of the sign can be resolved by a careful 
inspection.
\end{thebibliography}
\end{document}